\documentclass[10pt, conference]{IEEEtran}
\IEEEoverridecommandlockouts

\usepackage{xurl}
\usepackage[hidelinks]{hyperref}
\usepackage{cite}
\usepackage{amsmath,amssymb,amsfonts}
\usepackage{algorithmic}
\usepackage{graphicx}
\usepackage{textcomp}
\usepackage{xcolor}
\usepackage{enumitem}
\usepackage{subcaption}
\usepackage{balance}
\usepackage{booktabs}

\def\BibTeX{{\rm B\kern-.05em{\sc i\kern-.025em b}\kern-.08em
    T\kern-.1667em\lower.7ex\hbox{E}\kern-.125emX}}

\newcommand{\para}[1]{\noindent\textbf{\textit{#1}:}}

\begin{document}

\title{Towards Reproducible Evaluation of Distributed Quantum Circuit Partitioning Algorithms
\thanks{This work is supported in part by an NSF grant \#2426940.}}

\author{\IEEEauthorblockN{Javier Vela-Tambo}
\IEEEauthorblockA{\textit{Worcester Polytechnic Institute}\\
Worcester, MA, USA \\
jvela@wpi.edu}
\and
\IEEEauthorblockN{Davud Azizov}
\IEEEauthorblockA{\textit{Worcester Polytechnic Institute}\\
Worcester, MA, USA \\
dazizov@wpi.edu}
\and
\IEEEauthorblockN{Tian Guo}
\IEEEauthorblockA{\textit{Worcester Polytechnic Institute}\\
Worcester, MA, USA \\
tian@wpi.edu}
}

\maketitle

\begin{abstract}
Distributed Quantum Computing (DQC) addresses the physical scaling limitations of monolithic quantum processors by networking modular Quantum Processing Units (QPUs). Efficient execution of quantum algorithms on DQC architectures requires compiling them across QPUs while minimizing inter-QPU communication bottlenecks, primarily through circuit partitioning. However, current evaluations of state-of-the-art partitioning heuristics focus primarily on the total entanglement cost of the partitions, failing to capture the broader structural and temporal overheads introduced by distributed network constraints. This paper addresses this evaluation gap by applying established monolithic benchmarking metrics to partitioned distributed circuits to quantify the performance impact of network constraints. Using an open-source, automated evaluation pipeline, we systematically assess diverse partitioning algorithms across standardized workloads and quantum network topologies. Our empirical results reveal that partitioning algorithms with comparable entanglement costs can still introduce drastically different physical execution penalties. By exposing these hidden trade-offs, such as severe increases in circuit depth and substantial reductions in gate density, this study demonstrates that comprehensive circuit-level metrics are essential for guiding the future design of DQC compilers.
\end{abstract}


\section{Introduction}
\label{sec:introduction}

The scaling of monolithic Noisy Intermediate-Scale Quantum (NISQ) processors is constrained by physical limitations, such as decoherence, cross-talk, and processor topology~\cite{barralReviewDistributedQuantum2025, caleffiDistributedQuantumComputing2024}. Distributed Quantum Computing (DQC) offers a viable solution to this scaling challenge by networking modular Quantum Processing Units (QPUs) together to work as a unified computational system~\cite{barralReviewDistributedQuantum2025, angARQUINArchitecturesMultinode2024, mainDistributedQuantumComputing2025}. However, operating a DQC system introduces hardware and networking challenges, including the need for robust quantum memories, precise clock synchronization between remote nodes, and entanglement distillation to mitigate environmental noise~\cite{barralReviewDistributedQuantum2025, caleffiDistributedQuantumComputing2024, angARQUINArchitecturesMultinode2024}. Most notably, DQC systems suffer from a severe communication bottleneck. Because inter-QPU communication relies on the generation and distribution of fragile entanglement~\cite{bennettTeleportingUnknownQuantum1993,eisertOptimalLocalImplementation2000}, remote operations can be two to three orders of magnitude slower and noisier than local routing~\cite{angARQUINArchitecturesMultinode2024}.

\vfill\null
\newpage

Because of this communication bottleneck, compiling quantum algorithms for execution on a DQC architecture is non-trivial~\cite{caleffiDistributedQuantumComputing2024, angARQUINArchitecturesMultinode2024}. To execute these circuits, compilers must solve the \textit{circuit partitioning problem}: the task of mapping a monolithic quantum circuit onto the networked QPUs while minimizing the inter-QPU communication overhead caused by remote operations~\cite{barralReviewDistributedQuantum2025, caleffiDistributedQuantumComputing2024, yimsiriwattanaDistributedQuantumComputing2004}. While evaluations of current partitioning algorithms focus on the total entanglement cost~\cite{andres-martinezDistributingCircuitsHeterogeneous2024, bakerTimeslicedQuantumCircuit2020, burtMultilevelFrameworkPartitioning2026}, comprehensive circuit-level metrics remain confined to monolithic benchmarks~\cite{liQASMBenchLowLevelQuantum2023, tomeshSupermarQScalableQuantum2022}. Consequently, current DQC evaluations fail to capture the structural and temporal overheads introduced by distributed network constraints, leaving a significant evaluation gap.

Addressing this evaluation gap is the primary contribution of this work. Specifically, we evaluate partitioning methods by applying established circuit features directly to the \textit{distributed circuits}. To ensure our study is systematic and reproducible, we utilize a diverse collection of standardized circuit workloads drawn from established benchmark suites and DQC literature~\cite{liQASMBenchLowLevelQuantum2023, burtMultilevelFrameworkPartitioning2026}. To support reproducibility, we open-source our complete evaluation pipeline, making our codebase, benchmarks, and partitioned circuits publicly accessible\footnote{\textcolor{magenta}{\url{https://github.com/cake-lab/dqc-evaluation/}}}. Using this approach, we can accurately characterize the structural overheads generated by DQC partitioning algorithms, such as the inflation of gate count introduced by non-local teleportation primitives. By evaluating how these penalties behave across varying circuit sizes, workloads, and network topologies, our framework exposes critical physical execution bottlenecks.

Rather than proposing new partitioning heuristics, we focus on evaluation methodology. We aim to integrate DQC into the broader ecosystem of quantum software evaluation and benchmarking by applying established metrics to distributed circuits. Our results show that partitioning algorithms with comparable entanglement costs can still produce very different structural penalties, such as expanded circuit depth and lower gate density caused by forcing qubits to remain inactive during network routing. These findings demonstrate that structure-aware metrics are necessary for evaluating DQC compilers and for guiding future partitioning algorithm design.

\vfill\null
\newpage

This paper makes three primary contributions:
\begin{itemize}[leftmargin=.12in,topsep=4pt]
    \item \textbf{Methodological Evaluation:} We adapt monolithic benchmarking metrics to partitioned circuits, exposing physical execution trade-offs masked by entanglement costs.
    \item \textbf{Open-Source Pipeline:} We release an automated, reproducible benchmarking framework combining state-of-the-art partitioners, standardized quantum workloads, and configurable network topologies.
    \item \textbf{Empirical Analysis:} We quantify distribution overheads, demonstrating that heuristics optimizing purely for e-bits can severely inflate circuit depth and degrade gate density.
\end{itemize}

The remainder of this paper is organized as follows. We begin by establishing the fundamental mechanics of DQC communication and reviewing current partitioning strategies and benchmarking methodologies (\S\ref{sec:background_related_work}). We then detail our evaluation design, defining the extended structural metrics alongside the selected circuit workloads and physical network constraints (\S\ref{sec:evaluation_design_methodology}). This methodology provides the foundation for an empirical analysis of how these partitioning methods perform across varying algorithmic structures and network topologies (\S\ref{sec:results_analysis}). Finally, we summarize our key findings and discuss future directions (\S\ref{sec:conclusion}).

\begin{figure*}[!htbp]
\centering
\includegraphics{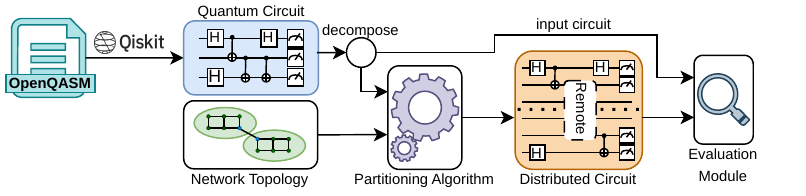}
\vspace{1mm}
\caption{The evaluation pipeline. Monolithic OpenQASM quantum circuits are ingested and decomposed using Qiskit. The circuits are then distributed to the specified network topology using the selected partitioning algorithms. The original monolithic circuit and the distributed circuit, which contains explicit remote gate instructions, are profiled by our evaluation module to extract the desired metrics. The pipeline is controlled by a YAML configuration file to ensure the reproducibility of the experiments.}
\label{fig:design}
\vspace{1mm}
\end{figure*}

\section{Background \& Related Work}
\label{sec:background_related_work}

\subsection{Distributed Quantum Computing Primitives}

To satisfy the connectivity requirements of a partitioned circuit, non-local operations must be performed between remote QPUs. These non-local operations rely on the consumption of shared entanglement, specifically Einstein-Podolsky-Rosen (EPR) pairs, commonly referred to as \textit{e-bits}. Remote operations are generally executed via distinct types of teleportation protocols. The first is \textit{quantum state teleportation}, which uses local operations and classical communication to physically relocate the quantum state of a data qubit from one QPU to another, consuming an e-bit in the process~\cite{bennettTeleportingUnknownQuantum1993}. 

Alternatively, non-local controlled operations can be executed between QPUs without moving the control qubit via \textit{quantum gate teleportation}, referred to as the EJPP protocol~\cite{eisertOptimalLocalImplementation2000}. This is achieved using the \textit{cat-entangler} and \textit{cat-disentangler} primitives~\cite{yimsiriwattanaGeneralizedGHZStates2004}. Furthermore, gate teleportation is a flexible procedure that allows for \textit{extended gate teleportation}, or gate packing. Because the control line can be distributed and maintained via a \textit{cat-like state}, multiple non-local gates sharing the same control qubit can be packed together to reuse a single e-bit, directly reducing the entanglement cost of the partitioned circuit~\cite{eisertOptimalLocalImplementation2000, yimsiriwattanaGeneralizedGHZStates2004}.

However, the execution of state or gate teleportation requires additional local operations, intermediate measurements, and classical communication, which add multiple layers of depth to the distributed circuit. Additionally, the communication overhead introduces significant latency, forcing qubits to idle during remote operations and increasing their susceptibility to phase decoherence, which degrades the fidelity of the computation~\cite{barralReviewDistributedQuantum2025, caleffiDistributedQuantumComputing2024}.

\subsection{Circuit Partitioning Algorithms}

To map monolithic quantum circuits onto distributed architectures, circuit partitioning is typically addressed through classical heuristic algorithms designed to minimize inter-QPU communication (the entanglement cost). Because the partitioning problem and the underlying qubit mapping are known to be NP-hard~\cite{itoAlgorithmicTheoryQubit2023}, exact optimal solutions become computationally intractable as circuit depth and qubit count grow. State-of-the-art partitioning methods rely on distinct approximations to tackle this bottleneck.

Baker~\textit{et al.} introduced a \textit{fine-grained}, \textit{time-sliced partitioning} (FGP) approach that reaches a single static assignment by mapping quantum circuits one layer (time-slice) at a time~\cite{bakerTimeslicedQuantumCircuit2020}. Their method generates a sequence of local interaction graphs and uses look-ahead weights to ensure that teleportations also consider future slices. As a result of this approach, all non-local operations are ultimately covered by state teleportation.

Adopting a global distribution strategy, Andrés-Martínez~\textit{et al.} leverage \textit{hypergraph partitioning} to model multi-qubit gate groups as hyperedges. This framework directly matches edge cuts to e-bit costs rather than simply minimizing the non-local gate count. This approach is highly motivated by extended gate teleportation, focusing on packing sequences of gates into the same EJPP teleportation protocol to minimize communication overhead globally across the entire circuit~\cite{andres-martinezAutomatedDistributionQuantum2019, andres-martinezDistributingCircuitsHeterogeneous2024}.

Burt~\textit{et al.} proposed a multilevel framework that addresses the limitations of static interaction graphs. Their approach utilizes temporal coarsening to group gates into a sequence of hierarchical graphs~\cite{burtMultilevelFrameworkPartitioning2026}. By applying heuristics such as the Fiduccia-Mattheyses algorithm across these coarsened levels, their method jointly optimizes both state and gate teleportation.

While these three frameworks form the primary graph-based baselines evaluated in this study, alternative optimization techniques have also emerged. For instance, recent works explore simulated annealing and Tabu search to optimize qubit assignments across heterogeneous topologies~\cite{zhouOptimizingCompilationDistributed2025, sundaramDistributionQuantumCircuits2022}, as well as leveraging Deep Reinforcement Learning (DRL) to optimize qubit routing under specific distributed hardware constraints~\cite{pastorCircuitPartitioningMultiCore2024, promponasCompilerDistributedQuantum2025, russoOptimizingQubitAssignment2025}.

\subsection{Quantum Circuit Benchmarking}

To evaluate quantum software and optimization frameworks, researchers rely on established benchmark suites, including QASMBench and SupermarQ~\cite{liQASMBenchLowLevelQuantum2023, tomeshSupermarQScalableQuantum2022}. QASMBench provides a collection of low-level, commonly used quantum routines in the OpenQASM representation and introduces metrics, such as \textit{gate density}, \textit{retention lifespan}, and \textit{entanglement variance}, to characterize execution efficiency and susceptibility to hardware noise~\cite{liQASMBenchLowLevelQuantum2023}. Similarly, SupermarQ offers a scalable, hardware-agnostic suite that quantifies quantum programs using application-level feature vectors, such as \textit{critical-depth}, \textit{parallelism}, and \textit{liveness}~\cite{tomeshSupermarQScalableQuantum2022}.

Although these suites provide established architectural profiles, their use remains confined to monolithic quantum circuits~\cite{liQASMBenchLowLevelQuantum2023, tomeshSupermarQScalableQuantum2022}. At the same time, the state-of-the-art partitioning algorithms~\cite{bakerTimeslicedQuantumCircuit2020, andres-martinezDistributingCircuitsHeterogeneous2024, burtMultilevelFrameworkPartitioning2026} are evaluated almost exclusively on the total entanglement (e-bit) cost of the partition and classical execution runtime. Our work explicitly bridges this gap by porting these structural metrics directly into the distributed circuit partitioning domain.

\section{Evaluation Design \& Methodology}
\label{sec:evaluation_design_methodology}

To systematically evaluate the performance of the selected partitioning methods, we developed an automated evaluation pipeline, illustrated in \figurename~\ref{fig:design}. The pipeline's execution is controlled via YAML configuration files to ensure repeatable experiments. First, monolithic OpenQASM quantum circuits~\cite{crossOpenQuantumAssembly2017} (\S\ref{sec:design:circuits}) are ingested and decomposed using Qiskit~\cite{qiskit2024}. These circuits are then passed to the partitioning algorithm (\S\ref{sec:design:algs}), which distributes the operations onto the target hardware's network topology (\S\ref{sec:design:networks}). The output is a partitioned, distributed circuit containing explicit remote gate instructions. Finally, both the monolithic and the distributed circuit are profiled by our evaluation module to extract the desired metrics (\S\ref{sec:design:metrics}).

\subsection{Circuit Benchmarks \& Workloads}
\label{sec:design:circuits}

Our study uses workloads from the QASMBench suite~\cite{liQASMBenchLowLevelQuantum2023} alongside a custom dataset of quantum circuits generated according to the methodology outlined by Burt~\textit{et al.}~\cite{burtMultilevelFrameworkPartitioning2026} to ensure a comprehensive evaluation across a wide spectrum of algorithmic structures. Since the exact circuits are not available, we regenerated them following their methodology and saved the configurations to ensure our work is fully reproducible. Generated workloads span sizes from 16 to 96 qubits, with 10 distinct instances per size for randomized circuit types.

\para{QASMBench} To evaluate on real-world quantum applications, we include the circuits from the established QASMBench suite~\cite{liQASMBenchLowLevelQuantum2023}. We utilize a diverse set of circuits spanning the \textit{small} (2--10 qubits), \textit{medium} (11--27 qubits), and \textit{large} (28--433 qubits) categories. QASMBench enables the evaluation of how partitioning methods handle varying degrees of structural complexity and spatial locality in practical, gate-level workloads, preventing the evaluation biases often introduced by purely random circuits~\cite{liQASMBenchLowLevelQuantum2023}.

\para{Quantum Fourier Transform (QFT)} The QFT features a highly dense, all-to-all qubit interaction pattern~\cite{barralReviewDistributedQuantum2025, sundaramEfficientDistributionQuantum2021}. This makes it a worst-case scenario for distributed computing, as partitioning it forces heavy non-local routing with an $O(n^2)$ communication overhead~\cite{yimsiriwattanaGeneralizedGHZStates2004}. Despite this, its regular structure forms large distributable packets of gates, making it an essential benchmark for evaluating gate teleportation strategies~\cite{burtGeneralisedCircuitPartitioning2024}.

\para{Quantum Approximate Optimization Algorithm (QAOA)} QAOA is a variational quantum-classical algorithm comprising parameterized single- and two-qubit gates~\cite{farhiQuantumApproximateOptimization2014}. Similar to the QFT, QAOA circuits permit the grouping of gates for teleportation~\cite{burtMultilevelFrameworkPartitioning2026}. In contrast to the rigid density of the QFT, QAOA gate interactions are naturally commuting, which allows compilers to freely reorder operations~\cite{tomeshSupermarQScalableQuantum2022, cuomoOptimizedCompilerDistributed2023} and makes QAOA an excellent benchmark for evaluating how well partitioning methods exploit circuit structure~\cite{bakerTimeslicedQuantumCircuit2020, tomeshSupermarQScalableQuantum2022}. We evaluate QAOA circuits generated with a 50\% edge probability.

\para{Quantum Volume (QV)} Quantum Volume circuits are a randomized metric originally designed to benchmark the overall performance of a quantum computer~\cite{crossValidatingQuantumComputers2019}. Structurally, a QV circuit consists of alternating blocks of repeated single- and two-qubit Haar-random gate patterns interleaved with random permutations of qubits. QV circuits represent a crucial middle ground between highly structured algorithms and fully random circuits~\cite{burtGeneralisedCircuitPartitioning2024}. Due to their randomized structure, QV circuits lack large distributable packets of gates. Consequently, partitioning algorithms reliant on gate teleportation typically perform poorly, whereas those employing state teleportation remain efficient~\cite{burtGeneralisedCircuitPartitioning2024}. For all generated QV circuits, we set the number of circuit layers equal to the number of qubits.

\para{Controlled-Phase Fraction (CP)} CP-fraction circuits are a generalization of CZ-fraction circuits~\cite{sundaramEfficientDistributionQuantum2021}, consisting of layers where each qubit either undergoes a single-qubit gate or is randomly paired with another qubit via a two-qubit gate based on a specific probability~\cite{burtGeneralisedCircuitPartitioning2024}. Because their structure is entirely non-uniform, CP-fraction circuits make it inefficient for partitioning methods to constrain themselves to state teleportation~\cite{burtGeneralisedCircuitPartitioning2024}. For our evaluation, we generated circuits at gate densities of 0.3, 0.5, 0.7, and 0.9.

\begin{table}[t]
\centering
\caption{Breakdown of execution outcomes (\%) for each partitioning algorithm across all experiments.}
\label{tab:partitioner_outcomes}
\resizebox{\linewidth}{!}{%
\begin{tabular}{lcccc}
\toprule
\textbf{\shortstack{Partitioning\\Method}} & 
\textbf{\shortstack{Success\\(\%)}} & 
\textbf{\shortstack{Failure\\(\%)}} & 
\textbf{\shortstack{Partitioning\\Timeout (\%)}} & 
\textbf{\shortstack{Postprocessing\\Timeout (\%)}} \\
\midrule
FGP-rOEE   & 81.48 & 1.27 & 17.25 &  0.00 \\
MLFM-R     & 90.41 & 7.88 &  1.68 &  0.03 \\
Pytket P   & 33.67 & 1.65 &  7.27 & 57.40 \\
Pytket PE  & 34.28 & 1.78 & 32.78 & 31.16 \\
Pytket ESD & 32.81 & 1.91 & 35.87 & 29.42 \\
\bottomrule
\end{tabular}%
}
\end{table}

\subsection{Partitioning Algorithms}
\label{sec:design:algs}

We evaluate three distinct families of state-of-the-art partitioning algorithms, using the open-source software frameworks provided by their respective authors.

\begin{itemize}[leftmargin=.12in,topsep=4pt]
    \item \textbf{MLFM}: The Multilevel Fiduccia-Mattheyses with Recursive coarsening (MLFM-R) algorithm, proposed by Burt~\textit{et al.}~\cite{burtMultilevelFrameworkPartitioning2026} and implemented as part of DISQCO~\cite{felix-burtFelixburtDISQCO2026}.
    \item \textbf{FGP}: The Fine-Grained Partitioning relaxed Overall Extreme Exchange (FGP-rOEE) algorithm, proposed by Baker~\textit{et al.}~\cite{bakerTimeslicedQuantumCircuit2020} and also implemented within DISQCO~\cite{felix-burtFelixburtDISQCO2026}.
    \item \textbf{Pytket-DQC}: Three algorithm variants proposed by Andrés-Martínez~\textit{et al.}: Partitioning (P), Partitioning with Embedding (PE), and Cover Embedding Steiner Detached (ESD)~\cite{andres-martinezDistributingCircuitsHeterogeneous2024}, implemented in the Pytket-DQC library~\cite{QuantinuumPytketdqc2026}.
\end{itemize}

Because these tools use different internal representations, we developed a unified wrapper for the DISQCO and Pytket-DQC frameworks. The wrapper ingests the Qiskit-decomposed circuits, executes each partitioning algorithm\footnote{The wrapper enforces a 600-second timeout for all partitioning executions.}, and exports the distributed circuits into a unified format with the gates and measurements required for remote operation.

\subsection{Network Topologies}
\label{sec:design:networks}

To measure routing overhead across varying connectivity, we evaluate three distinct inter-QPU topologies in four configurations: a 4-QPU \textit{linear} network, in which QPUs are arranged in a 1D line and each node is connected to its immediate neighbors; a 4-QPU \textit{grid} network, in which QPUs are arranged in a 2D matrix with horizontal and vertical neighbor connectivity; and 2-QPU and 4-QPU \textit{fully connected} networks, in which every QPU shares a direct link with all others.

In each configuration, the qubit capacity per QPU is $\lceil n_q / k \rceil + 1$, where $n_q$ is the total number of circuit qubits and $k$ is the number of QPUs. This allocation provides an auxiliary communication qubit in each QPU to facilitate non-local teleportation primitives. Furthermore, intra-QPU connectivity is assumed to be fully connected, allowing any two qubits within the same module to interact without internal routing penalties. An abstraction layer translates these topologies into framework-specific network definitions from DISQCO~\cite{felix-burtFelixburtDISQCO2026} and Pytket-DQC~\cite{QuantinuumPytketdqc2026}.

\begin{figure}[t]
\centering
\includegraphics{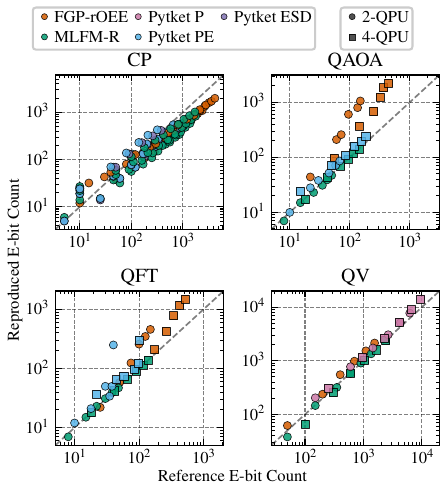}
\caption{Replication analysis of entanglement cost (measured in e-bits) across the CP, QAOA, QFT, and QV benchmarks. Each point compares our reproduced e-bit count against the approximated reference value reported in prior literature~\cite{burtMultilevelFrameworkPartitioning2026}.}
\label{fig:results:comparison}
\end{figure}

\subsection{Evaluation Metrics}
\label{sec:design:metrics}

Our DQC compilation evaluation combines direct partitioning costs with circuit-level metrics. Direct costs include \textit{e-bit count}, which measures the number of remote entanglement pairs consumed by non-local operations, and \textit{partitioning execution time}, which captures classical compiler overhead. We also report standard circuit properties: \textit{width}, the number of active qubits; \textit{depth}, the number of circuit time steps; and \textit{gate count}, the total number of operations. Because remote communication can inflate these values, they indicate whether the distributed circuit remains executable within the physical limits of a target DQC architecture.

Structural penalties are captured using QASMBench and SupermarQ features~\cite{liQASMBenchLowLevelQuantum2023,tomeshSupermarQScalableQuantum2022}: \textit{gate density} measures how fully the circuit spacetime is occupied, revealing idle slots caused by remote communication; \textit{retention lifespan} and \textit{liveness} measure how long qubits remain active and exposed to noise; \textit{parallelism} captures execution concurrency; \textit{entanglement ratio} measures the fraction of entangling gates; and \textit{entanglement variance} captures whether two-qubit interactions are concentrated into hotspots that partitioners can map onto QPU nodes to reduce e-bit cost.

Finally, we analyze these metrics as absolute values, ratios, and deltas ($\Delta$) to quantify structural changes between the monolithic and distributed circuits. All metrics are extracted by parsing Qiskit's Directed Acyclic Graph (DAG) circuit representation~\cite{qiskit2024}.

\begin{figure}[t]
\centering
\includegraphics{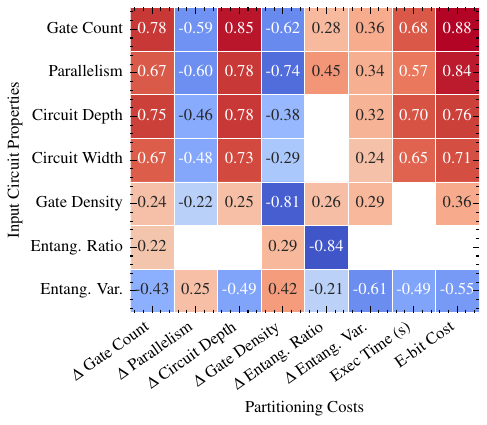}
\caption{Spearman correlation matrix mapping input quantum circuit properties against DQC partitioning cost and structural penalties. Darker shades indicate stronger correlations (red for positive, blue for negative). Missing entries denote correlations that are not statistically significant ($p > 0.05$) or fall below the strength threshold ($|r| < 0.20$).}
\label{fig:results:correlation}
\end{figure}

\section{Results \& Analysis}
\label{sec:results_analysis}

We quantify the overheads introduced by DQC partitioning across quantum workloads and network topologies using the metrics in \S\ref{sec:design:metrics}. To ensure a fair comparison, we report only circuit instances that completed successfully within the timeout for all compared partitioning methods and, where applicable, all network topologies.

Table~\ref{tab:partitioner_outcomes} summarizes the execution status breakdown. Across all partitioners, MLFM-R and FGP-rOEE achieve the highest completion rates (90.4\% and 81.5\%), with FGP-rOEE's remaining overhead driven almost entirely by partitioning timeouts (17.3\%). In contrast, Pytket variants drop to $\sim$33--34\% success, limited primarily by postprocessing timeouts (up to 57.4\% for Pytket~P) and partitioning timeouts (up to 35.9\% for Pytket~ESD), while failure rates remain low ($\le 7.9\%$) across all methods.

\subsection{Replication Analysis}

To validate our evaluation pipeline, we compare our partitioning e-bit costs with those of Burt~\textit{et al.}~\cite{burtMultilevelFrameworkPartitioning2026} for 2-QPU and 4-QPU fully connected inter-QPU networks. \figurename~\ref{fig:results:comparison} reports reproduced e-bit counts for MLFM-R, FGP-rOEE, and the Pytket variants across CP, QAOA, QFT, and QV benchmarks, using reference values approximated from the published figures. The results closely follow the $y=x$ parity line, confirming pipeline correctness; minor scatter is expected from stochastic partitioning and randomized benchmark circuits that may differ from the reference study. FGP-rOEE generally shows higher e-bit costs for QAOA and QFT.

\begin{figure}[!t]
\centering
\includegraphics{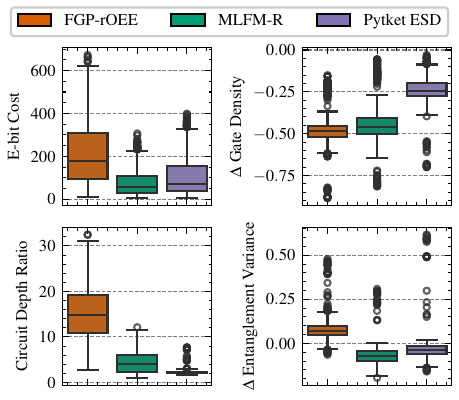}
\caption{Multidimensional evaluation of DQC partitioning overheads. The grouped box plots contrast e-bit cost, circuit depth ratio, $\Delta$ gate density, and $\Delta$ entanglement variance across partitioning methods.}
\label{fig:results:cost_metrics}
\end{figure}

\subsection{Predictors of Partitioning Overhead}
\label{sec:results:correlation}

We identify drivers of compilation degradation through Spearman rank correlations ($r$) between monolithic circuit properties, partitioning costs, and structural penalties ($\Delta$). We remove redundant variables from highly correlated groups and show only statistically significant relationships ($p \le 0.05$) with $|r| \ge 0.20$ in \figurename~\ref{fig:results:correlation}.

\para{Primary Predictors of Communication Overhead} Communication overhead (e-bit cost) is driven by workload scale and concurrency. Its strongest predictors are gate count ($r=0.88$) and parallelism ($r=0.84$): as circuit concurrency grows, more interacting qubits span QPU boundaries, consuming more e-bits for non-local gates. The only metric that significantly reduces communication overhead is entanglement variance ($r=-0.55$), confirming that algorithms localize dense interaction hotspots to single nodes and minimize e-bit cost.

\para{Structural Trade-offs and Penalties} Because remote operations inflate structural metrics, we characterize how partitioning alters output circuits and affects noisy-hardware behavior. Non-local teleportation primitives force qubits to wait, reducing utilization and gate density ($r=-0.81$). High input parallelism and gate counts strongly correlate with expanded circuit depth ($r=0.78$ and $r=0.85$), since concurrent operations must be serialized across the network, extending maximum qubit lifespan and increasing noise exposure. Circuits with high initial entanglement ratio also suffer a severe decrease in entanglement ratio ($r=-0.84$).

\para{Algorithmic Execution Bottlenecks} Compilation execution time scales strongly with input-circuit size, driven by depth ($r=0.70$), width ($r=0.65$), and gate count ($r=0.68$). Gate density has no statistically significant correlation with execution time, indicating that compiler latency depends on circuit scale rather than spatial crowding of active gate slots.

\subsection{The Cost of Distribution and Network Connectivity}

\figurename~\ref{fig:results:cost_metrics} aggregates the primary metrics identified in \S\ref{sec:results:correlation}, showing the multidimensional costs of distribution across partitioning algorithms. For example, MLFM-R and Pytket-ESD have comparable median e-bit counts, yet MLFM-R inflates circuit depth by a significantly higher ratio. Although all algorithms reduce gate density, degradation varies widely and is most pronounced in FGP-rOEE. Thus, minimizing e-bit count is necessary but insufficient, as it obscures severe structural and temporal overheads.

\begin{figure}[!t]
\centering
\includegraphics{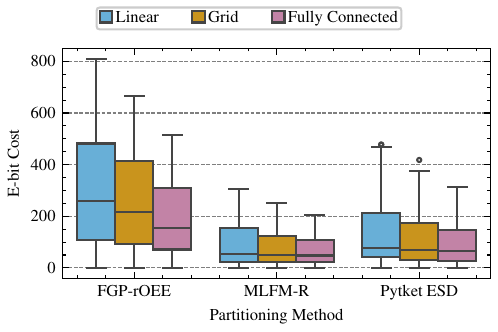}
\caption{Comparison of DQC partitioning e-bit cost across different 4-QPU network topologies (linear, grid, and fully connected) and partitioning methods.}
\label{fig:results:network_topology}
\end{figure}

\figurename~\ref{fig:results:network_topology} compares e-bit costs across 4-QPU linear, grid, and fully connected topologies. As expected, restricted connectivity penalizes partitioning: moving from fully connected to linear topology forces multi-hop routing and significantly inflates e-bit consumption.

\section{Conclusion}
\label{sec:conclusion}

This study shows that evaluating DQC partitioning algorithms solely by total entanglement (e-bit) cost is insufficient. Our results demonstrate that partitioning algorithms with similar e-bit costs can still introduce different structural penalties, including expanded circuit depth, reduced gate density, and longer qubit lifetimes. These findings highlight the need for structure-aware evaluation metrics that better capture the executability of distributed quantum circuits. This study has two main limitations. First, the results are not disaggregated by individual benchmark and circuit instance, which could reveal more detailed workload-specific trends. Second, some completed partitioning runs could not be included because postprocessing into the common representation timed out. In future work, we plan to expand this framework by integrating noisy-hardware simulation to directly quantify execution fidelity, adding partitioning algorithms, and extending network-topology models to reflect the precise physical constraints of real-world QPU architectures.

\vfill\null
\newpage

\balance
\bibliography{references}

\begin{thebibliography}{10}
\providecommand{\url}[1]{#1}
\csname url@samestyle\endcsname
\providecommand{\newblock}{\relax}
\providecommand{\bibinfo}[2]{#2}
\providecommand{\BIBentrySTDinterwordspacing}{\spaceskip=0pt\relax}
\providecommand{\BIBentryALTinterwordstretchfactor}{4}
\providecommand{\BIBentryALTinterwordspacing}{\spaceskip=\fontdimen2\font plus
\BIBentryALTinterwordstretchfactor\fontdimen3\font minus \fontdimen4\font\relax}
\providecommand{\BIBforeignlanguage}[2]{{%
\expandafter\ifx\csname l@#1\endcsname\relax
\typeout{** WARNING: IEEEtran.bst: No hyphenation pattern has been}%
\typeout{** loaded for the language `#1'. Using the pattern for}%
\typeout{** the default language instead.}%
\else
\language=\csname l@#1\endcsname
\fi
#2}}
\providecommand{\BIBdecl}{\relax}
\BIBdecl

\bibitem{barralReviewDistributedQuantum2025}
\BIBentryALTinterwordspacing
D.~Barral, F.~J. Cardama, G.~Díaz-Camacho, D.~Faílde, I.~F. Llovo, M.~Mussa-Juane, J.~Vázquez-Pérez, J.~Villasuso, C.~Piñeiro, N.~Costas, J.~C. Pichel, T.~F. Pena, and A.~Gómez, ``Review of {Distributed} {Quantum} {Computing}: {From} single {QPU} to {High} {Performance} {Quantum} {Computing},'' \emph{Computer Science Review}, vol.~57, p. 100747, Aug. 2025. [Online]. Available: \url{https://doi.org/10.1016/j.cosrev.2025.100747}
\BIBentrySTDinterwordspacing

\bibitem{caleffiDistributedQuantumComputing2024}
\BIBentryALTinterwordspacing
M.~Caleffi, M.~Amoretti, D.~Ferrari, J.~Illiano, A.~Manzalini, and A.~S. Cacciapuoti, ``Distributed quantum computing: {A} survey,'' \emph{Computer Networks}, vol. 254, p. 110672, Dec. 2024. [Online]. Available: \url{https://doi.org/10.1016/j.comnet.2024.110672}
\BIBentrySTDinterwordspacing

\bibitem{angARQUINArchitecturesMultinode2024}
\BIBentryALTinterwordspacing
J.~Ang, G.~Carini, Y.~Chen, I.~Chuang, M.~Demarco, S.~Economou, A.~Eickbusch, A.~Faraon, K.-M. Fu, S.~Girvin, M.~Hatridge, A.~Houck, P.~Hilaire, K.~Krsulich, A.~Li, C.~Liu, Y.~Liu, M.~Martonosi, D.~McKay, J.~Misewich, M.~Ritter, R.~Schoelkopf, S.~Stein, S.~Sussman, H.~Tang, W.~Tang, T.~Tomesh, N.~Tubman, C.~Wang, N.~Wiebe, Y.~Yao, D.~Yost, and Y.~Zhou, ``{ARQUIN}: {Architectures} for {Multinode} {Superconducting} {Quantum} {Computers},'' \emph{ACM Transactions on Quantum Computing}, vol.~5, no.~3, pp. 19:1--19:59, Sep. 2024. [Online]. Available: \url{https://doi.org/10.1145/3674151}
\BIBentrySTDinterwordspacing

\bibitem{mainDistributedQuantumComputing2025}
\BIBentryALTinterwordspacing
D.~Main, P.~Drmota, D.~P. Nadlinger, E.~M. Ainley, A.~Agrawal, B.~C. Nichol, R.~Srinivas, G.~Araneda, and D.~M. Lucas, ``\BIBforeignlanguage{en}{Distributed quantum computing across an optical network link},'' \emph{\BIBforeignlanguage{en}{Nature}}, vol. 638, no. 8050, pp. 383--388, Feb. 2025. [Online]. Available: \url{https://doi.org/10.1038/s41586-024-08404-x}
\BIBentrySTDinterwordspacing

\bibitem{bennettTeleportingUnknownQuantum1993}
\BIBentryALTinterwordspacing
C.~H. Bennett, G.~Brassard, C.~Crépeau, R.~Jozsa, A.~Peres, and W.~K. Wootters, ``Teleporting an unknown quantum state via dual classical and {Einstein}-{Podolsky}-{Rosen} channels,'' \emph{Physical Review Letters}, vol.~70, no.~13, pp. 1895--1899, Mar. 1993. [Online]. Available: \url{https://doi.org/10.1103/PhysRevLett.70.1895}
\BIBentrySTDinterwordspacing

\bibitem{eisertOptimalLocalImplementation2000}
\BIBentryALTinterwordspacing
J.~Eisert, K.~Jacobs, P.~Papadopoulos, and M.~B. Plenio, ``Optimal local implementation of nonlocal quantum gates,'' \emph{Physical Review A}, vol.~62, no.~5, p. 052317, Oct. 2000. [Online]. Available: \url{https://doi.org/10.1103/PhysRevA.62.052317}
\BIBentrySTDinterwordspacing

\bibitem{yimsiriwattanaDistributedQuantumComputing2004}
\BIBentryALTinterwordspacing
A.~Yimsiriwattana and S.~J.~L. Jr, ``\BIBforeignlanguage{en}{Distributed quantum computing: a distributed {Shor} algorithm},'' in \emph{\BIBforeignlanguage{en}{Quantum {Information} and {Computation} {II}}}, vol. 5436.\hskip 1em plus 0.5em minus 0.4em\relax SPIE, Aug. 2004, p. 360. [Online]. Available: \url{https://doi.org/10.1117/12.546504}
\BIBentrySTDinterwordspacing

\bibitem{andres-martinezDistributingCircuitsHeterogeneous2024}
\BIBentryALTinterwordspacing
P.~Andres-Martinez, T.~Forrer, D.~Mills, J.-Y. Wu, L.~Henaut, K.~Yamamoto, M.~Murao, and R.~Duncan, ``\BIBforeignlanguage{en}{Distributing circuits over heterogeneous, modular quantum computing network architectures},'' \emph{\BIBforeignlanguage{en}{Quantum Science and Technology}}, vol.~9, no.~4, p. 045021, Aug. 2024. [Online]. Available: \url{https://doi.org/10.1088/2058-9565/ad6734}
\BIBentrySTDinterwordspacing

\bibitem{bakerTimeslicedQuantumCircuit2020}
\BIBentryALTinterwordspacing
J.~M. Baker, C.~Duckering, A.~Hoover, and F.~T. Chong, ``Time-sliced quantum circuit partitioning for modular architectures,'' in \emph{Proceedings of the 17th {ACM} {International} {Conference} on {Computing} {Frontiers}}, ser. {CF} '20.\hskip 1em plus 0.5em minus 0.4em\relax Association for Computing Machinery, May 2020, pp. 98--107. [Online]. Available: \url{https://doi.org/10.1145/3387902.3392617}
\BIBentrySTDinterwordspacing

\bibitem{burtMultilevelFrameworkPartitioning2026}
\BIBentryALTinterwordspacing
F.~Burt, K.-C. Chen, and K.~K. Leung, ``\BIBforeignlanguage{en-GB}{A {Multilevel} {Framework} for {Partitioning} {Quantum} {Circuits}},'' \emph{\BIBforeignlanguage{en-GB}{Quantum}}, vol.~10, p. 1984, Jan. 2026. [Online]. Available: \url{https://doi.org/10.22331/q-2026-01-22-1984}
\BIBentrySTDinterwordspacing

\bibitem{liQASMBenchLowLevelQuantum2023}
\BIBentryALTinterwordspacing
A.~Li, S.~Stein, S.~Krishnamoorthy, and J.~Ang, ``{QASMBench}: {A} {Low}-{Level} {Quantum} {Benchmark} {Suite} for {NISQ} {Evaluation} and {Simulation},'' \emph{ACM Transactions on Quantum Computing}, vol.~4, no.~2, pp. 10:1--10:26, Feb. 2023. [Online]. Available: \url{https://doi.org/10.1145/3550488}
\BIBentrySTDinterwordspacing

\bibitem{tomeshSupermarQScalableQuantum2022}
\BIBentryALTinterwordspacing
T.~Tomesh, P.~Gokhale, V.~Omole, G.~S. Ravi, K.~N. Smith, J.~Viszlai, X.-C. Wu, N.~Hardavellas, M.~R. Martonosi, and F.~T. Chong, ``{SupermarQ}: {A} {Scalable} {Quantum} {Benchmark} {Suite},'' in \emph{2022 {IEEE} {International} {Symposium} on {High}-{Performance} {Computer} {Architecture} ({HPCA})}, Apr. 2022, pp. 587--603. [Online]. Available: \url{https://doi.org/10.1109/HPCA53966.2022.00050}
\BIBentrySTDinterwordspacing

\bibitem{yimsiriwattanaGeneralizedGHZStates2004}
\BIBentryALTinterwordspacing
A.~Yimsiriwattana and S.~J. Lomonaco~Jr., ``Generalized {GHZ} {States} and {Distributed} {Quantum} {Computing},'' Mar. 2004. [Online]. Available: \url{https://doi.org/10.48550/arXiv.quant-ph/0402148}
\BIBentrySTDinterwordspacing

\bibitem{itoAlgorithmicTheoryQubit2023}
T.~Ito, N.~Kakimura, N.~Kamiyama, Y.~Kobayashi, and Y.~Okamoto, ``\BIBforeignlanguage{en}{Algorithmic {Theory} of {Qubit} {Routing}},'' in \emph{\BIBforeignlanguage{en}{Algorithms and {Data} {Structures}}}, P.~Morin and S.~Suri, Eds.\hskip 1em plus 0.5em minus 0.4em\relax Springer Nature Switzerland, 2023, pp. 533--546.

\bibitem{andres-martinezAutomatedDistributionQuantum2019}
\BIBentryALTinterwordspacing
P.~Andrés-Martínez and C.~Heunen, ``Automated distribution of quantum circuits via hypergraph partitioning,'' \emph{Physical Review A}, vol. 100, no.~3, p. 032308, Sep. 2019. [Online]. Available: \url{https://doi.org/10.1103/PhysRevA.100.032308}
\BIBentrySTDinterwordspacing

\bibitem{zhouOptimizingCompilationDistributed2025}
\BIBentryALTinterwordspacing
R.~Zhou, J.~Cheng, Y.~Gan, J.~Liu, and C.~Qian, ``Optimizing {Compilation} for {Distributed} {Quantum} {Computing} via {Clustering} and {Annealing},'' in \emph{2025 {IEEE} {International} {Conference} on {Quantum} {Computing} and {Engineering} ({QCE})}, vol.~01, Aug. 2025, pp. 1312--1318. [Online]. Available: \url{https://doi.org/10.1109/QCE65121.2025.00146}
\BIBentrySTDinterwordspacing

\bibitem{sundaramDistributionQuantumCircuits2022}
\BIBentryALTinterwordspacing
R.~G. Sundaram, H.~Gupta, and C.~R. Ramakrishnan, ``Distribution of {Quantum} {Circuits} {Over} {General} {Quantum} {Networks},'' in \emph{2022 {IEEE} {International} {Conference} on {Quantum} {Computing} and {Engineering} ({QCE})}, Sep. 2022, pp. 415--425. [Online]. Available: \url{https://doi.org/10.1109/QCE53715.2022.00063}
\BIBentrySTDinterwordspacing

\bibitem{pastorCircuitPartitioningMultiCore2024}
\BIBentryALTinterwordspacing
A.~Pastor, P.~Escofet, S.~Ben~Rached, E.~Alarcón, P.~Barlet-Ros, and S.~Abadal, ``Circuit {Partitioning} for {Multi}-{Core} {Quantum} {Architectures} with {Deep} {Reinforcement} {Learning},'' in \emph{2024 {IEEE} {International} {Symposium} on {Circuits} and {Systems} ({ISCAS})}, May 2024, pp. 1--5. [Online]. Available: \url{https://doi.org/10.1109/ISCAS58744.2024.10557956}
\BIBentrySTDinterwordspacing

\bibitem{promponasCompilerDistributedQuantum2025}
\BIBentryALTinterwordspacing
P.~Promponas, A.~Mudvari, L.~D. Chiesa, P.~Polakos, L.~Samuel, and L.~Tassiulas, ``Compiler for {Distributed} {Quantum} {Computing}: {A} {Reinforcement} {Learning} {Approach},'' in \emph{{ICC} 2025 - {IEEE} {International} {Conference} on {Communications}}, Jun. 2025, pp. 4615--4621. [Online]. Available: \url{https://doi.org/10.1109/ICC52391.2025.11161115}
\BIBentrySTDinterwordspacing

\bibitem{russoOptimizingQubitAssignment2025}
\BIBentryALTinterwordspacing
E.~Russo, M.~Palesi, D.~Patti, G.~Ascia, and V.~Catania, ``Optimizing {Qubit} {Assignment} in {Modular} {Quantum} {Systems} via {Attention}-{Based} {Deep} {Reinforcement} {Learning},'' in \emph{2025 {Design}, {Automation} \& {Test} in {Europe} {Conference} ({DATE})}, Mar. 2025, pp. 1--7. [Online]. Available: \url{https://doi.org/10.23919/DATE64628.2025.10992725}
\BIBentrySTDinterwordspacing

\bibitem{crossOpenQuantumAssembly2017}
\BIBentryALTinterwordspacing
A.~W. Cross, L.~S. Bishop, J.~A. Smolin, and J.~M. Gambetta, ``Open {Quantum} {Assembly} {Language},'' Jul. 2017. [Online]. Available: \url{https://doi.org/10.48550/arXiv.1707.03429}
\BIBentrySTDinterwordspacing

\bibitem{qiskit2024}
\BIBentryALTinterwordspacing
A.~{Javadi-Abhari}, M.~Treinish, K.~Krsulich, C.~J. Wood, J.~Lishman, J.~Gacon, S.~Martiel, P.~D. Nation, L.~S. Bishop, A.~W. Cross, B.~R. Johnson, and J.~M. Gambetta, ``Quantum computing with {{Qiskit}},'' Jun. 2024. [Online]. Available: \url{https://doi.org/10.48550/arXiv.2405.08810}
\BIBentrySTDinterwordspacing

\bibitem{sundaramEfficientDistributionQuantum2021}
\BIBentryALTinterwordspacing
R.~G. Sundaram, H.~Gupta, and C.~R. Ramakrishnan, ``Efficient {Distribution} of {Quantum} {Circuits},'' in \emph{35th {International} {Symposium} on {Distributed} {Computing} ({DISC} 2021)}, ser. Leibniz {International} {Proceedings} in {Informatics} ({LIPIcs}), S.~Gilbert, Ed., vol. 209.\hskip 1em plus 0.5em minus 0.4em\relax Schloss Dagstuhl – Leibniz-Zentrum für Informatik, 2021, pp. 41:1--41:20. [Online]. Available: \url{https://doi.org/10.4230/LIPIcs.DISC.2021.41}
\BIBentrySTDinterwordspacing

\bibitem{burtGeneralisedCircuitPartitioning2024}
\BIBentryALTinterwordspacing
F.~Burt, K.-C. Chen, and K.~K. Leung, ``Generalised {Circuit} {Partitioning} for {Distributed} {Quantum} {Computing},'' in \emph{2024 {IEEE} {International} {Conference} on {Quantum} {Computing} and {Engineering} ({QCE})}, vol.~02, Sep. 2024, pp. 173--178. [Online]. Available: \url{https://doi.org/10.1109/QCE60285.2024.10273}
\BIBentrySTDinterwordspacing

\bibitem{farhiQuantumApproximateOptimization2014}
\BIBentryALTinterwordspacing
E.~Farhi, J.~Goldstone, and S.~Gutmann, ``A {Quantum} {Approximate} {Optimization} {Algorithm},'' Nov. 2014. [Online]. Available: \url{https://doi.org/10.48550/arXiv.1411.4028}
\BIBentrySTDinterwordspacing

\bibitem{cuomoOptimizedCompilerDistributed2023}
\BIBentryALTinterwordspacing
D.~Cuomo, M.~Caleffi, K.~Krsulich, F.~Tramonto, G.~Agliardi, E.~Prati, and A.~S. Cacciapuoti, ``\BIBforeignlanguage{en}{Optimized {Compiler} for {Distributed} {Quantum} {Computing}},'' \emph{\BIBforeignlanguage{en}{ACM Transactions on Quantum Computing}}, vol.~4, no.~2, pp. 1--29, Jun. 2023. [Online]. Available: \url{https://doi.org/10.1145/3579367}
\BIBentrySTDinterwordspacing

\bibitem{crossValidatingQuantumComputers2019}
\BIBentryALTinterwordspacing
A.~W. Cross, L.~S. Bishop, S.~Sheldon, P.~D. Nation, and J.~M. Gambetta, ``Validating quantum computers using randomized model circuits,'' \emph{Physical Review A}, vol. 100, no.~3, p. 032328, Sep. 2019. [Online]. Available: \url{https://doi.org/10.1103/PhysRevA.100.032328}
\BIBentrySTDinterwordspacing

\bibitem{felix-burtFelixburtDISQCO2026}
\BIBentryALTinterwordspacing
felix burt, ``felix-burt/{DISQCO},'' Feb. 2026. [Online]. Available: \url{https://github.com/felix-burt/DISQCO}
\BIBentrySTDinterwordspacing

\bibitem{QuantinuumPytketdqc2026}
\BIBentryALTinterwordspacing
Quantinuum, ``Quantinuum/pytket-dqc,'' Jun. 2026. [Online]. Available: \url{https://github.com/Quantinuum/pytket-dqc}
\BIBentrySTDinterwordspacing

\end{thebibliography}

\end{document}